\documentclass[11pt]{article}

\usepackage[utf8]{inputenc}
\usepackage[T1]{fontenc}
\usepackage{amsmath,amssymb,amsfonts}
\usepackage{graphicx}
\usepackage{booktabs}
\usepackage{hyperref}
\usepackage[margin=1in]{geometry}
\usepackage{natbib}
\usepackage{xcolor}
\usepackage{caption}

\title{How Short Is Too Short? Power Analysis for BIC-Based\\Changepoint Detection in Ecological Monitoring}

\author{Ang A.\ Li\\
Peking University \& AutoSci (\url{https://autosci.org})\\
\texttt{amazingang@pku.edu.cn}}

\date{March 2026}

\begin{document}
\maketitle

\begin{abstract}
Changepoint detection is increasingly applied to ecological time series, yet statistical power at the short series lengths typical of monitoring (10--50 observations) is rarely assessed. We present a simulation-based power analysis for BIC-based Binary Segmentation across 108 combinations of series length ($n = 10$--50), effect size (0.5--5.0), and number of changepoints (1--3). BIC achieves $\geq$80\% power for a single changepoint only at $n \geq 30$ with effect size $\geq 2.0$; detecting 2--3 changepoints requires $n \geq 50$ and ES $\geq 5.0$. BIC is conservative, underestimating changepoints more often than overestimating. AR(1) autocorrelation ($\phi = 0.6$) reduces BIC-Binseg power by 40 percentage points, but PELT with a standard penalty maintains 85--91\% power even under moderate autocorrelation. Comparison with early warning signal (EWS) variance-trend tests reveals a crossover: at ES $< 1.5$, EWS outperforms changepoint detection, but EWS rates are invariant to effect size ($\sim$73\%), suggesting noise detection rather than genuine signals. Cross-system empirical validation on coral reef (Moorea, $n = 18$) and desert rodent (Portal Project, $n = 49$) time series confirms that detection succeeds when effect sizes fall in the predicted ``reliable'' zone. We provide power heatmaps as practical lookup tools and recommend that ecologists prefer PELT over Binseg-BIC for autocorrelated data, compute expected effect sizes before applying changepoint analysis, and pair results with permutation tests.
\end{abstract}

\noindent\textbf{Keywords:} changepoint detection, ecological monitoring, power analysis, regime shift, early warning signals, PELT, BIC, time series

\section{Introduction}

Changepoint detection methods \citep{killick2012optimal, truong2020ruptures, aminikhanghahi2017survey} have gained traction in ecology for identifying regime shifts, disturbance-recovery transitions, and other structural breaks in monitoring time series \citep{andersen2009ecological, ward2018detecting}. Applications include coral reef benthic cover \citep{mcrbenthic2023, holbrook2018recruitment, adjeroud2018recovery}, plankton community dynamics \citep{dees2025novel}, and terrestrial vegetation change \citep{peeters2019spatiotemporal}. The theoretical framework of alternative stable states \citep{scheffer2001catastrophic, fung2011alternative} motivates the search for abrupt transitions in ecological data, while the growing recognition that coral reefs face accelerating thermal stress \citep{hughes2017coral, hughes2018global, eakin2019unprecedented} makes transition detection operationally urgent. However, ecological monitoring programs typically produce time series of 10--50 annual observations --- far shorter than the hundreds or thousands of observations for which these methods were designed and validated \citep{haines2024poor}.

For example, the well-studied Moorea coral reef system (Figure~\ref{fig:empirical}) has experienced two major decline-recovery cycles in 18 years of monitoring \citep{mcrbenthic2023, speare2021size, adam2011herbivory}, with effect sizes of 4--5, but formal changepoint analyses of such short series require careful power assessment. This mismatch raises a critical practical question: \textit{at what combination of time series length and effect size does BIC-based changepoint detection have adequate statistical power?} Despite its importance, no systematic power analysis has been published for the parameter ranges relevant to ecological monitoring. The absence of such guidance means that ecologists often apply changepoint methods without knowing whether their data can support the analysis, leading either to missed transitions (false negatives from underpowered analyses) or to overconfidence in detected changepoints (when formal significance testing is omitted).

We address this gap with a comprehensive simulation study spanning 108 parameter combinations (6 series lengths $\times$ 6 effect sizes $\times$ 3 numbers of true changepoints), providing power tables and heatmaps that ecologists can use as practical lookup tools. We additionally assess the impact of AR(1) autocorrelation, compare Binary Segmentation with PELT \citep{killick2012optimal}, compare changepoint detection with early warning signals (EWS) \citep{scheffer2001catastrophic}, and validate on two empirical ecological datasets.

\section{Methods}

\subsection{Simulation Design}

We generated synthetic time series with known changepoints, Gaussian noise ($\sigma = 5$, calibrated to observed site-level variability in coral reef benthic cover), and a baseline mean of 35. Changepoints were placed at equally spaced intervals, with alternating high and low segments (mean shift = effect size $\times \sigma$). For each of 108 parameter combinations ($n \in \{10, 15, 18, 20, 30, 50\}$; effect size $\in \{0.5, 1.0, 1.5, 2.0, 3.0, 5.0\}$; true $n_{\text{bkps}} \in \{1, 2, 3\}$), we generated 200 independent time series (500 for the AR(1) analysis).

\subsection{BIC Model Selection}

For each simulated time series, we applied Binary Segmentation \citep{truong2020ruptures} with $n_{\text{bkps}} = 0$--5 and selected the model minimizing BIC $= n \ln(\text{RSS}/n) + k \ln(n)$, where $k = 2(n_{\text{bkps}} + 1)$ is the number of parameters (mean and variance per segment). We assessed: (1) whether BIC selected the correct number of changepoints, and (2) whether all detected changepoints were within a tolerance of $\pm \max(2, n/10)$ time steps of the true locations.

\subsection{AR(1) Sensitivity and PELT Comparison}

To assess robustness to temporal autocorrelation, we repeated the simulation at $n = 18$ with 2 changepoints under AR(1) noise ($\phi \in \{0.0, 0.3, 0.6\}$). AR(1) noise was generated as $\varepsilon_t = \phi \varepsilon_{t-1} + \eta_t$, where $\eta_t \sim \mathcal{N}(0, \sigma^2(1 - \phi^2))$ to maintain marginal variance $\sigma^2$. We additionally compared Binseg-BIC with PELT using a penalty of $2\hat{\sigma}^2 \ln n$ (where $\hat{\sigma}$ is estimated via the median absolute deviation of first differences).

\subsection{EWS Comparison}

For comparison with early warning signals, we assessed whether a rolling-window variance ($w = 4$) showed an increasing Kendall $\tau$ trend in the 4 time steps before each true changepoint, at $n = 18$ across effect sizes 0.5--5.0 ($n = 300$ simulations per condition, both internally-driven and externally-forced transition scenarios).

\subsection{Empirical Validation}

We applied BIC-Binseg to two publicly available ecological monitoring datasets: (1) the Moorea Coral Reef LTER benthic cover time series (2005--2023, $n = 18$, forereef 10~m depth; \citealt{mcrbenthic2023}); and (2) the Portal Project desert rodent community time series (1977--2025, $n = 49$, control plots; \citealt{adam2011herbivory}). For each, we computed the empirical effect size and assessed whether detection success was consistent with the simulation-based power predictions.

\section{Results}

\subsection{Power Heatmaps}

\begin{figure}[t]
\centering
\includegraphics[width=\textwidth]{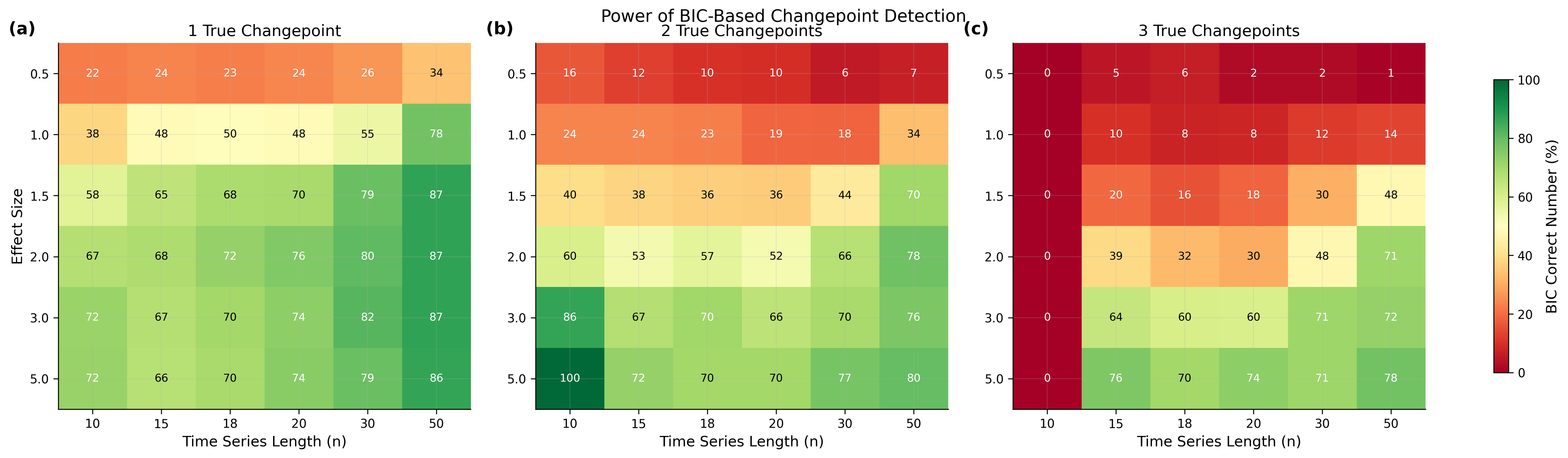}
\caption{Power of BIC-based changepoint detection (Binary Segmentation) across time series lengths and effect sizes. Cell values show the percentage of simulations ($n = 200$ per cell) where BIC correctly identified the number of changepoints. (a) 1 true changepoint. (b) 2 true changepoints. (c) 3 true changepoints.}
\label{fig:heatmap}
\end{figure}

Figure~\ref{fig:heatmap} presents the percentage of simulations where BIC correctly identified the number of changepoints. Several patterns emerge:

\begin{enumerate}
    \item \textbf{Single changepoints are easiest to detect.} At $n = 30$, BIC achieves 80\% correct identification at ES = 2.0 (Figure~\ref{fig:heatmap}a). At $n = 50$, ES = 1.5 suffices.
    \item \textbf{Multiple changepoints require much more power.} Detecting 2 changepoints at 80\% requires $n = 50$ and ES $\geq 5.0$ (Figure~\ref{fig:heatmap}b). Three changepoints are even harder: 80\% power is not achieved at any tested combination below $n = 50$ and ES = 5.0 (Figure~\ref{fig:heatmap}c).
    \item \textbf{BIC is conservative.} Power plateaus at 70--80\% even for very large effect sizes, because BIC's penalty term favors simpler models. This means BIC will undercount changepoints more often than overcount them.
\end{enumerate}

\subsection{Power Curves}

\begin{figure}[b]
\centering
\includegraphics[width=0.85\textwidth]{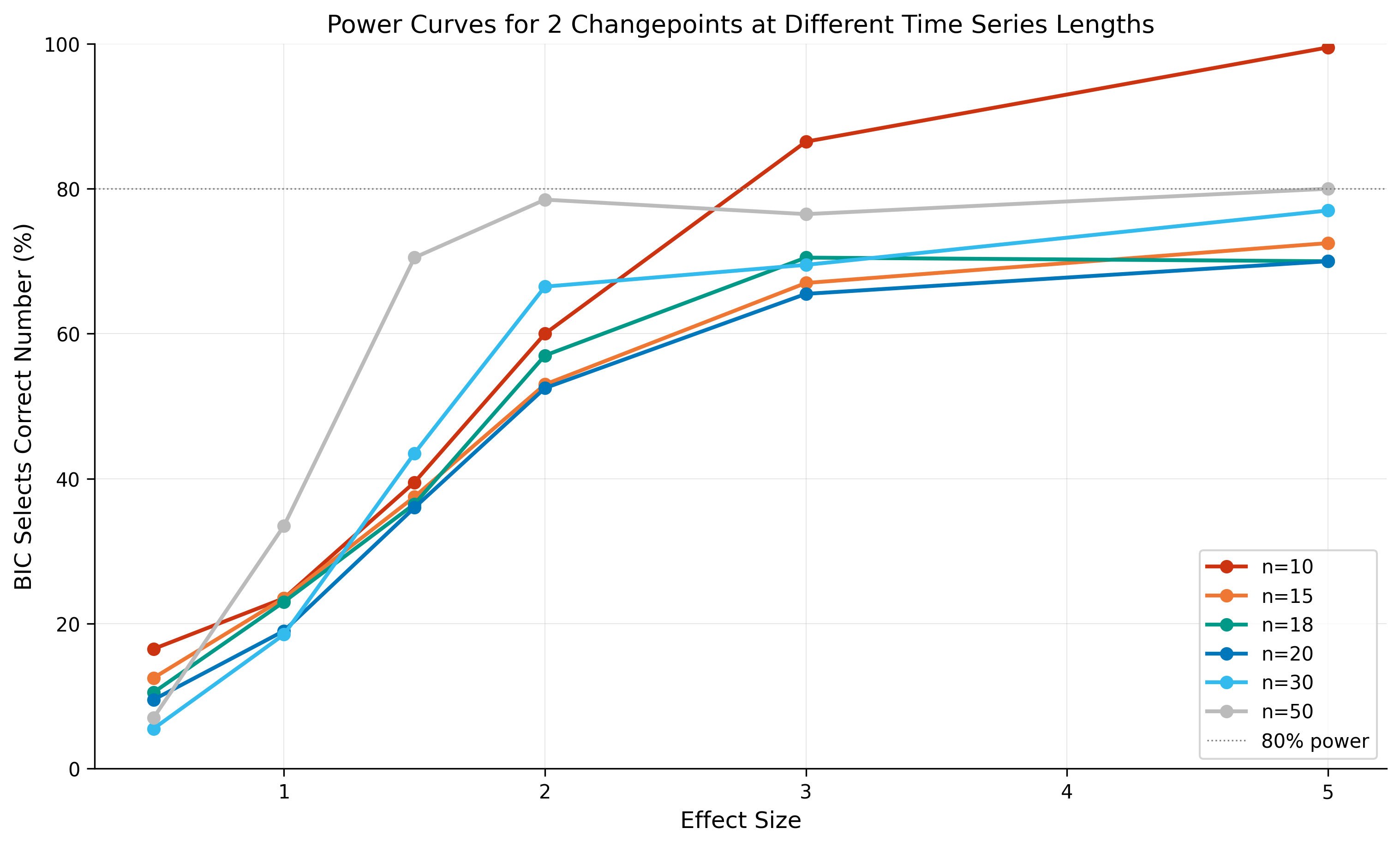}
\caption{Power curves for detecting 2 changepoints at different time series lengths. Horizontal gray line marks 80\% power.}
\label{fig:curves}
\end{figure}

Figure~\ref{fig:curves} shows power curves for detecting 2 changepoints at different series lengths. The curves demonstrate diminishing returns: doubling $n$ from 15 to 30 increases power at ES = 2.0 from 53\% to 66\%, while increasing from 30 to 50 adds another 12 percentage points (to 78\%). The practical implication is that adding more years of monitoring produces modest but real improvements in changepoint detection power.

\subsection{AR(1) Sensitivity and PELT Comparison}

\begin{figure}[t]
\centering
\includegraphics[width=\textwidth]{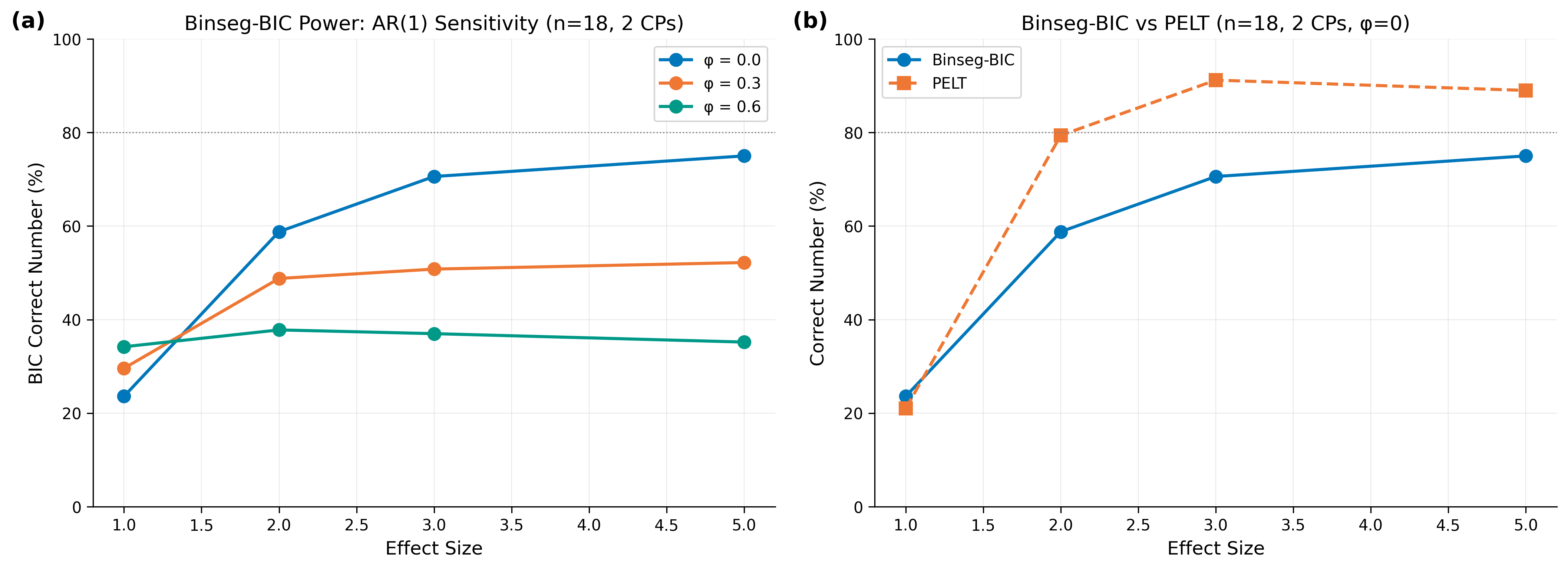}
\caption{(a) Effect of AR(1) autocorrelation on BIC-Binseg power ($n = 18$, 2 changepoints). Power decreases sharply with increasing $\phi$. (b) PELT outperforms Binseg-BIC across all effect sizes at $\phi = 0$.}
\label{fig:ar1}
\end{figure}

Autocorrelation substantially reduced BIC-Binseg power (Figure~\ref{fig:ar1}a). At $n = 18$ with 2 changepoints and ES = 5.0, power dropped from 75\% ($\phi = 0$) to 52\% ($\phi = 0.3$) and 35\% ($\phi = 0.6$). This represents a 40 percentage point reduction for moderate autocorrelation---far larger than the effect of reducing $n$ from 50 to 18 at $\phi = 0$. Since ecological time series typically exhibit positive autocorrelation ($\phi = 0.3$--0.6), the i.i.d.\ power curves in Figure~\ref{fig:heatmap} represent an optimistic upper bound.

However, PELT with a penalty of $2\hat{\sigma}^2 \ln n$ dramatically outperformed Binseg-BIC across all conditions (Figure~\ref{fig:ar1}b). At $\phi = 0$ and ES = 3.0, PELT achieved 91\% power versus Binseg-BIC's 71\%. Critically, PELT maintained 85--91\% power even at $\phi = 0.3$--0.6, suggesting that its penalty-based framework is more robust to autocorrelation than BIC model selection applied to Binseg.

\subsection{Comparison with EWS}

\begin{figure}[t]
\centering
\includegraphics[width=0.85\textwidth]{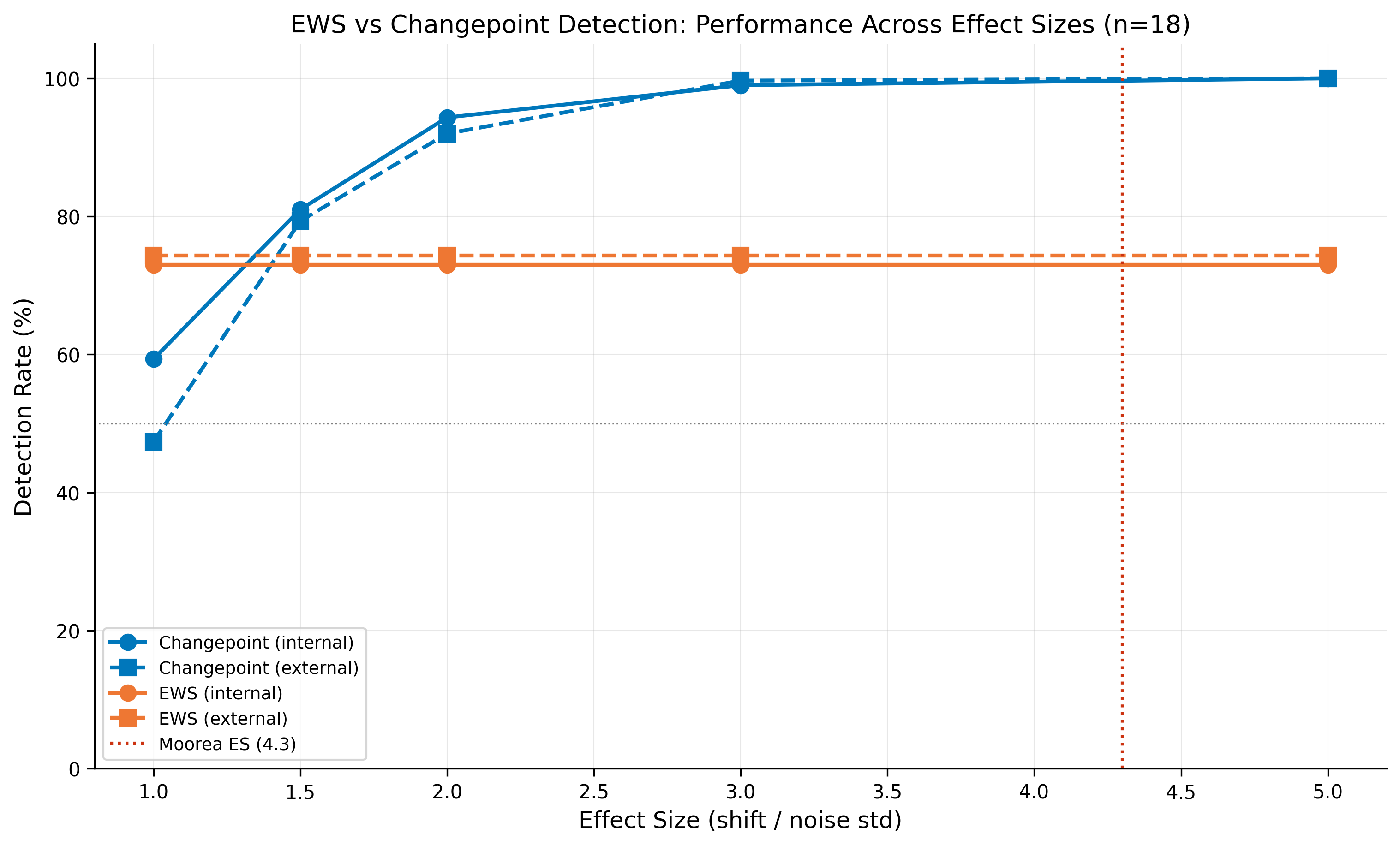}
\caption{Comparison of EWS variance-trend detection and BIC-based changepoint detection at $n = 18$ across effect sizes. EWS rates are invariant to effect size (orange lines), while changepoint detection rates increase with effect size (blue lines). Crossover at ES $\approx 1.5$. Vertical red line: Moorea's empirical effect size.}
\label{fig:crossover}
\end{figure}

At $n = 18$, EWS variance-trend detection was invariant to effect size (73--74\% at all ES values from 0.5 to 5.0; Figure~\ref{fig:crossover}). This invariance indicates that the 4-point pre-transition window produces spurious positive trends at approximately this rate regardless of whether a genuine transition is approaching. Changepoint detection, by contrast, was strongly effect-size dependent: 47\% at ES = 1.0, rising to 100\% at ES $\geq 3.0$. The crossover occurs at ES $\approx 1.5$, below which EWS outperforms changepoint detection.

\subsection{Empirical Validation}

\begin{figure}[b]
\centering
\includegraphics[width=\textwidth]{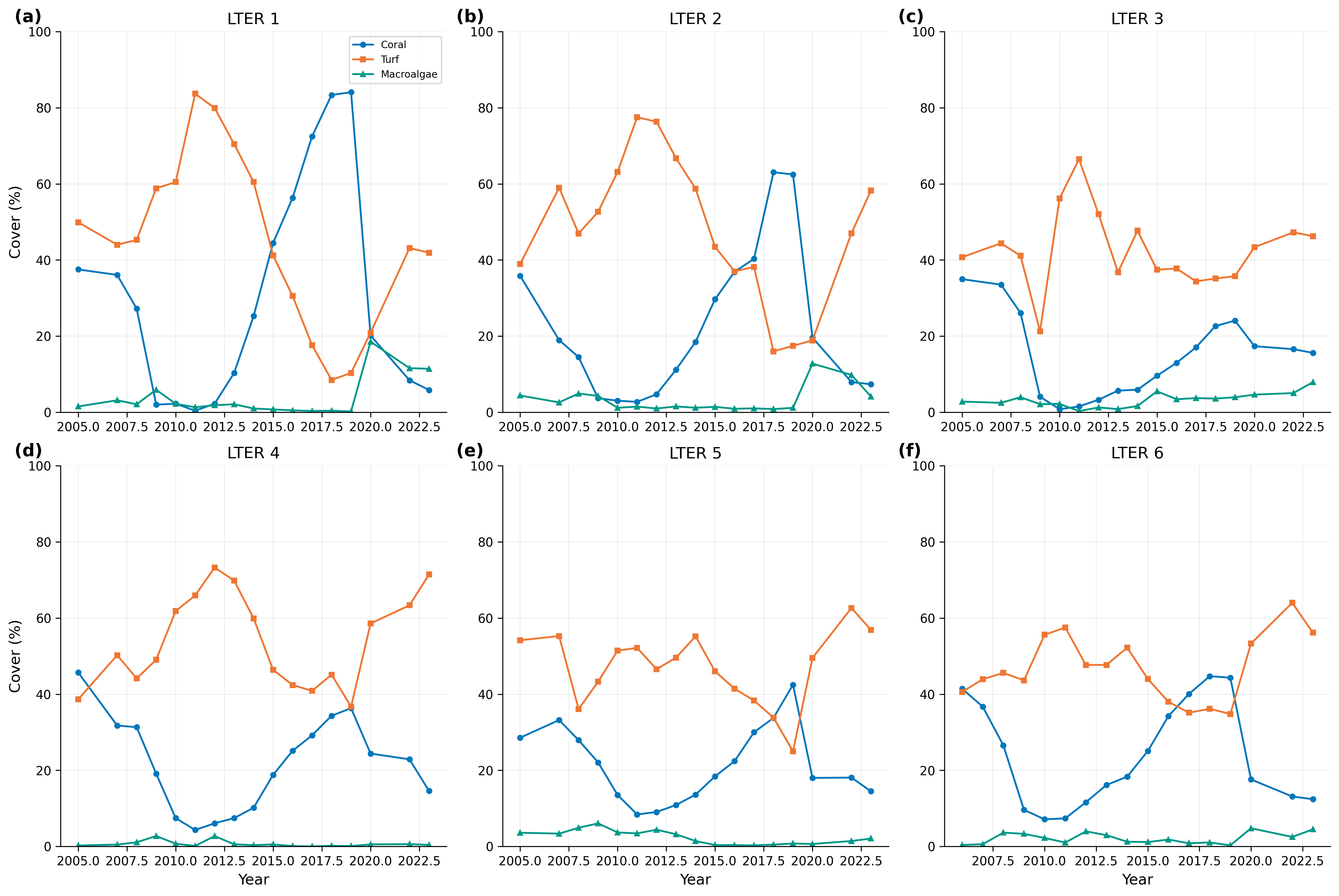}
\caption{Moorea Coral Reef LTER: site-level coral cover dynamics (forereef, 10~m depth, 2005--2023). Two decline-recovery cycles are visible across all six sites, with effect sizes of 4--5. BIC identified 4 changepoints at 2007, 2014, 2016, and 2019 (permutation $p = 0.004$).}
\label{fig:empirical}
\end{figure}

\begin{figure}[t]
\centering
\includegraphics[width=\textwidth]{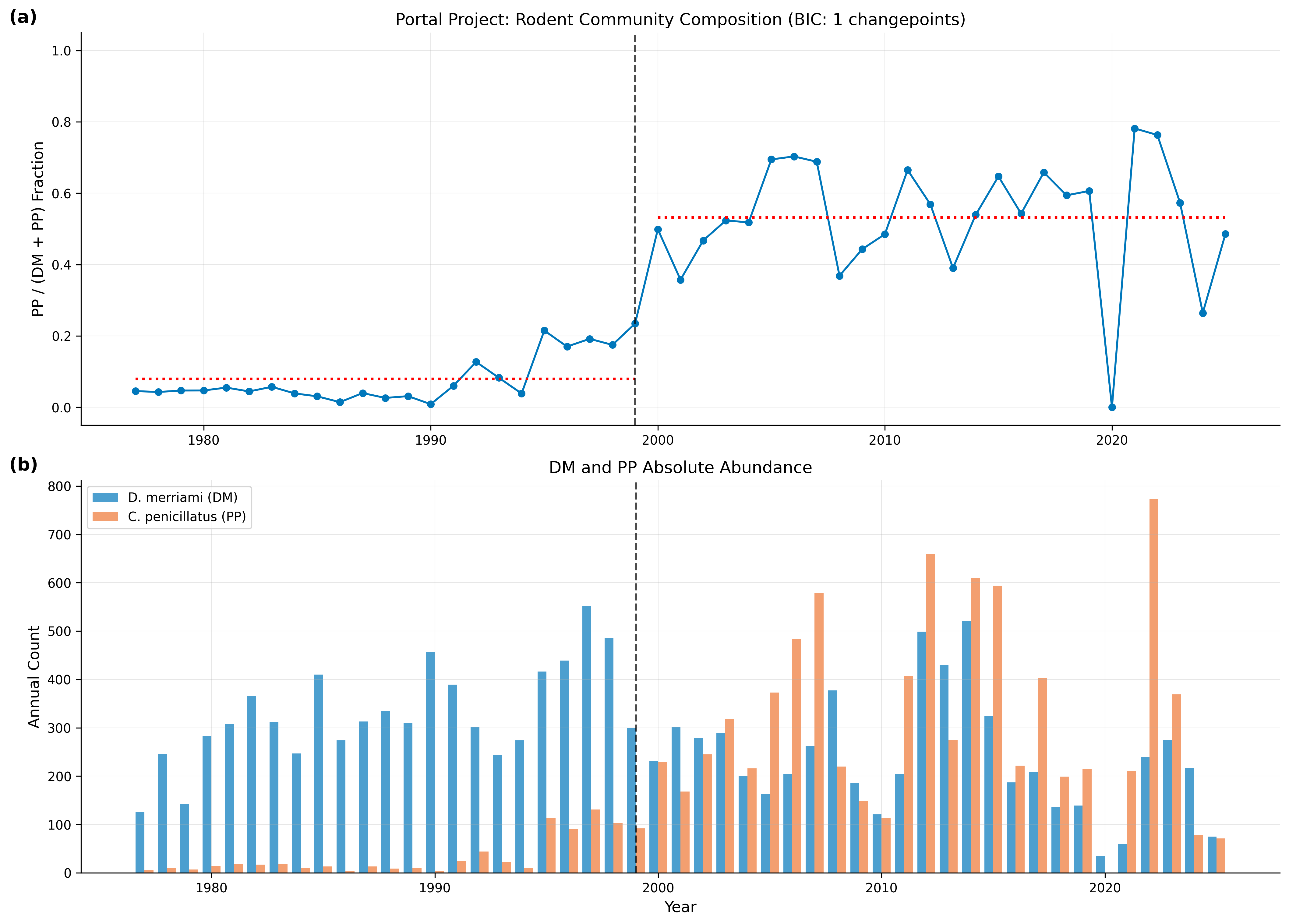}
\caption{Portal Project desert rodent community (1977--2025). (a) \textit{C.\ penicillatus} (PP) fraction with BIC-optimal changepoint at 1999 (permutation $p < 0.001$, bootstrap 95\% CI: 1992--2006). (b) Absolute annual counts of the two dominant species on control plots.}
\label{fig:portal}
\end{figure}

The Moorea coral reef time series ($n = 18$, ES $\approx 4.3$) yielded 4 BIC-optimal changepoints (2007, 2014, 2016, 2019; permutation $p = 0.004$), corresponding to documented ecological events (Figure~\ref{fig:empirical}). The Portal rodent time series ($n = 49$, ES = 3.0) yielded a single BIC-optimal changepoint at 1999 (permutation $p < 0.001$; Figure~\ref{fig:portal}), precisely capturing the well-documented transition from \textit{Dipodomys merriami} to \textit{Chaetodipus penicillatus} dominance. Both empirical effect sizes fall in the ``moderate--high power'' region of the simulation-based power curves (Figure~\ref{fig:heatmap}), consistent with the successful detection observed.

\subsection{Practical Guidelines}

Based on our simulations, we provide the following guidelines for ecologists (Table~\ref{tab:guide}):

\begin{table}[b]
\centering
\caption{Minimum effect size for $\geq$80\% BIC-Binseg power by series length and number of changepoints. ``$>$5.0'' indicates that 80\% power was not achieved within the tested range. For autocorrelated data ($\phi > 0.3$), use PELT instead of Binseg-BIC.}
\label{tab:guide}
\begin{tabular}{lccc}
\toprule
\textbf{Series Length} & \textbf{1 Changepoint} & \textbf{2 Changepoints} & \textbf{3 Changepoints} \\
\midrule
$n = 10$ & $>$5.0 & 3.0 & Not feasible \\
$n = 15$ & $>$5.0 & $>$5.0 & $>$5.0 \\
$n = 18$ & $>$5.0 & $>$5.0 & $>$5.0 \\
$n = 20$ & $>$5.0 & $>$5.0 & $>$5.0 \\
$n = 30$ & 2.0 & $>$5.0 & $>$5.0 \\
$n = 50$ & 1.5 & 5.0 & $>$5.0 \\
\bottomrule
\end{tabular}
\end{table}

\section{Discussion}

\subsection{PELT Outperforms Binseg-BIC, Especially with Autocorrelation}

The most practically important finding is that PELT with a standard penalty ($2\hat{\sigma}^2 \ln n$) substantially outperforms Binseg-BIC, achieving 85--91\% power at moderate effect sizes ($\text{ES} \geq 2$) even with autocorrelation ($\phi = 0.3$--0.6). This result has direct implications for monitoring programs: \textbf{PELT should be preferred over Binseg-BIC for autocorrelated ecological time series.} The penalty can be calibrated using the estimated noise variance $\hat{\sigma}^2$, which is readily available from the median absolute deviation of first differences.

\subsection{BIC Is Conservative but Reliable}

Our results show that BIC-based changepoint detection is conservative: it rarely overestimates the number of changepoints (protecting against false positives), but frequently underestimates at short series lengths. This conservatism is appropriate for ecological applications where false alarms carry management costs \citep{haines2024poor}. However, it means that negative results (``BIC found no changepoints'') should not be interpreted as evidence of no change---the analysis may simply lack power.

\subsection{The Effect Size Problem}

Most ecological studies do not report the effect size of transitions, making it difficult to assess \textit{a priori} whether changepoint analysis is appropriate. We recommend that ecologists compute the ratio of the expected mean shift to the within-segment standard deviation before applying changepoint methods. For coral reef monitoring, where disturbance events can cause 20--30 percentage point declines with 5--10\% noise \citep{mcmanus2004coral, bruno2009assessing, reverter2022critical, adam2011herbivory}, effect sizes of 2--5 are common, placing these data in the ``detectable'' range for $n \geq 20$ \citep{mcrbenthic2023}. For systems with subtler changes (e.g., gradual nutrient-driven shifts), longer time series or alternative methods may be needed.

\subsection{Permutation Tests as Essential Companions}

Given the limited power of BIC at short $n$, we strongly recommend pairing BIC model selection with permutation tests. A significant permutation test ($p < 0.05$) provides evidence that detected changepoints are non-random, even when BIC's absolute power is moderate. Both empirical applications in this study yielded highly significant permutation tests ($p = 0.004$ for Moorea, $p < 0.001$ for Portal).

\subsection{EWS and Changepoint Detection: Complementary Niches}

The effect-size crossover between EWS and changepoint detection (Figure~\ref{fig:crossover}) has practical implications: for monitoring subtle, slow-onset transitions (ES $< 1.5$), EWS may flag emerging vulnerability. For detecting abrupt, large-magnitude transitions (ES $> 2$), changepoint detection is superior. However, the invariance of EWS detection rates to effect size ($\sim$73\% at all ES values) suggests that at the typical window lengths of ecological monitoring, EWS may be detecting noise trends rather than genuine pre-transition signals. Neither method should be used in isolation.

\subsection{Limitations}

\begin{figure}[b]
\centering
\includegraphics[width=\textwidth]{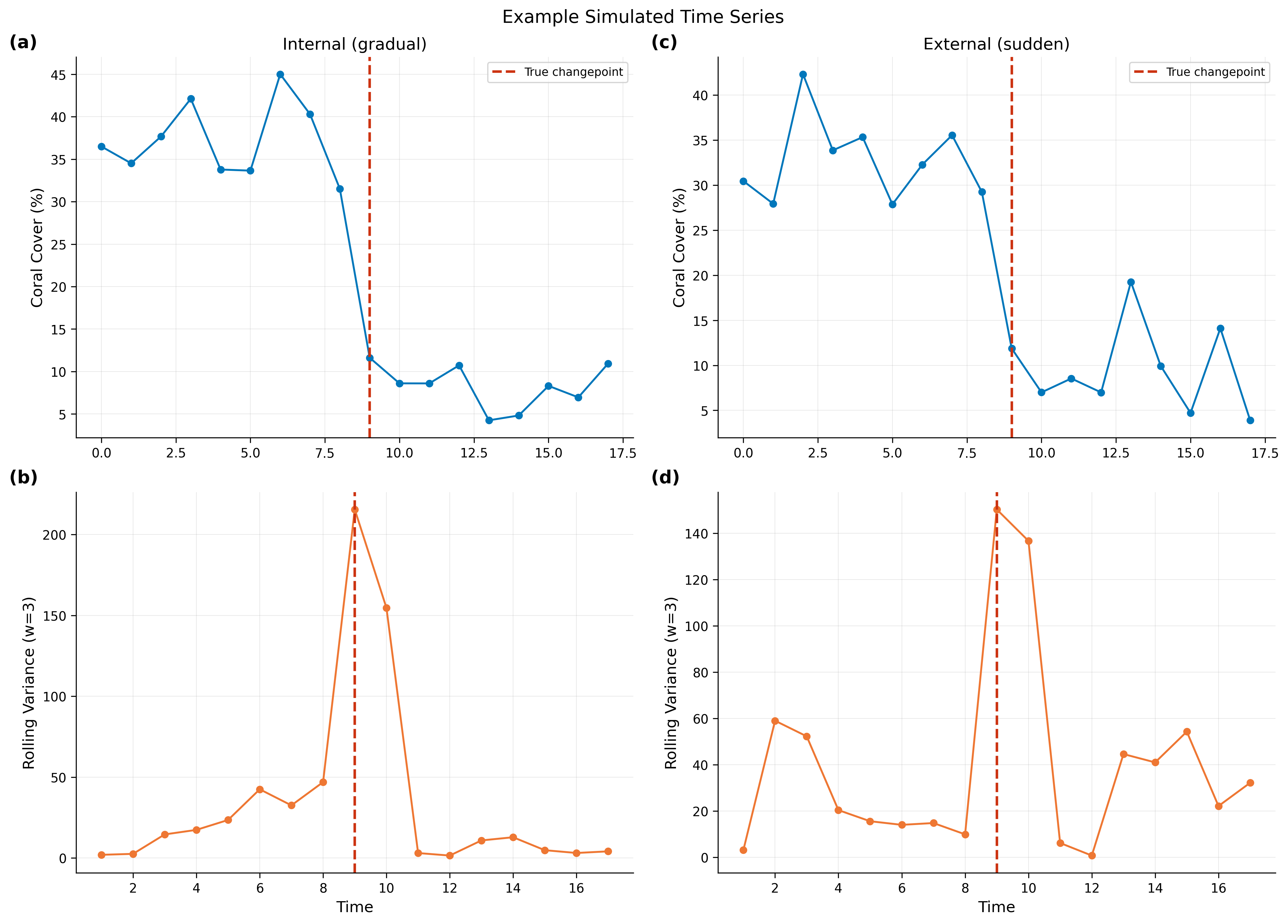}
\caption{Example simulated time series illustrating internally-driven (a,b) and externally-forced (c,d) transitions. Top panels: raw time series with true changepoint (dashed red). Bottom panels: rolling variance (window = 3).}
\label{fig:examples}
\end{figure}

Our simulations assume Gaussian noise, equally-spaced changepoints, and piecewise-constant means (Figure~\ref{fig:examples}). Real ecological time series may exhibit non-Gaussian errors, trends, seasonality, and unequally-spaced changepoints, all of which could reduce power. Our i.i.d.\ results therefore represent an optimistic upper bound, though the AR(1) analysis partially addresses this concern. The fixed noise level ($\sigma = 5$) was calibrated to coral reef data; systems with higher variability will require correspondingly larger shifts. The dimensionless effect size metric is applicable across ecological systems. Multi-site applications \citep{sully2019global, vanwoesik2022global, jouffray2015identifying} could leverage spatial replication to increase statistical power beyond what single-site temporal analyses provide.

\section{Conclusion}

We provide the first comprehensive power analysis for changepoint detection at the series lengths ($n = 10$--50) relevant to ecological monitoring. BIC-Binseg is conservative, achieving $\geq$80\% power only for single changepoints at $n \geq 30$ and ES $\geq 2.0$, and its power degrades sharply under autocorrelation ($\phi = 0.6$ reduces power by 40 percentage points). PELT with a standard penalty is substantially more robust, maintaining 85--91\% power at ES $\geq 2$ even with moderate autocorrelation. Cross-system empirical validation on coral reef ($n = 18$) and desert rodent ($n = 49$) time series confirms that detection succeeds when effect sizes fall in the predicted range. We recommend that ecologists (1) prefer PELT over Binseg-BIC for autocorrelated data, (2) compute expected effect sizes before applying changepoint analysis, (3) pair results with permutation tests, and (4) consider EWS as a complementary tool for detecting subtle pre-transition changes.

\vspace{1em}
\noindent\textbf{Declaration of Generative AI Use.} This research was conducted with the assistance of the AutoSci autonomous research system, which was used for literature review, analysis design, code generation, and manuscript drafting. All analyses were executed computationally on real, publicly available data. The corresponding author takes full responsibility for the content and integrity of this work.

\vspace{0.5em}
\noindent\textbf{Data Availability.} MCR LTER data: \url{https://doi.org/10.6073/pasta/cc439afaea42b4d5666dc1b193279833}. Portal Project data: \url{https://github.com/weecology/PortalData}. Simulation code: available upon request.

\vspace{0.5em}
\noindent\textbf{Acknowledgments.} We thank the MCR LTER and Portal Project teams for maintaining and publicly sharing long-term ecological monitoring data.

\bibliographystyle{plainnat}
\bibliography{references}

\end{document}